\documentclass[aps,prd,reprint,superscriptaddress]{revtex4-1}
\usepackage{amsmath,amssymb}
\usepackage[colorlinks=true,allcolors=blue]{hyperref}
\usepackage{dcolumn}
\usepackage{units}
\usepackage{graphicx}
\usepackage{float}
\usepackage{enumitem}
\usepackage{diagbox}

\allowdisplaybreaks[1]

\begin{document}

\title{Measuring test mass acceleration noise in \\ space-based gravitational wave astronomy}

\author{Giuseppe~Congedo}
\email{giuseppe.congedo@astro.ox.ac.uk}
\affiliation{Department of Physics, University of Oxford, Keble Road, Oxford OX1 3RH, UK}

\date{\today}

\begin{abstract}

The basic constituent of interferometric gravitational wave detectors -- the test mass to test mass interferometric link -- behaves as a differential dynamometer measuring effective differential forces, comprising an integrated measure of gravity curvature, inertial effects, as well as non-gravitational spurious forces. This last contribution is going to be characterised by the LISA Pathfinder mission, a technology precursor of future space-borne detectors like eLISA.
Changing the perspective from displacement to acceleration can benefit the data analysis of LISA Pathfinder and future detectors. The response in differential acceleration to gravitational waves is derived for a space-based detector's interferometric link. The acceleration formalism can also be integrated into time delay interferometry by building up the unequal-arm Michelson differential acceleration combination. The differential acceleration is nominally insensitive to the system free evolution dominating the slow displacement dynamics of low-frequency detectors. Working with acceleration also provides an effective way to subtract measured signals acting as systematics, including the actuation forces. Because of the strong similarity with the equations of motion, the optimal subtraction of systematic signals, known within some amplitude and time shift, with the focus on measuring the noise provides an effective way to solve the problem and marginalise over nuisance parameters. The $\mathcal{F}$-statistic, in widespread use throughout the gravitation waves community, is included in the method and suitably generalised to marginalise over linear parameters and noise at the same time. The method is applied to LPF simulator data and, thanks to its generality, can also be applied to the data reduction and analysis of future gravitational wave detectors.
 
\end{abstract}

\pacs{04.80.Nn, 04.80.-y, 04.30.-w, 07.05.Kf, 95.55.Ym, 95.75.-z}
%\keywords{}

\maketitle

% this is the main section

\section{Introduction}

The future space-based gravitational
wave (GW) detectors like eLISA \cite{elisa2013,amaro2012,elisa,bender1998}
will shed
new light on the low-frequency GW astrophysics, in the
frequency range $\unit[0.1-100]{mHz}$. At such frequencies,
space-based GW detectors are expected be sensitive to large
numbers of super-massive black hole binaries (see e.g.\
Ref.\;\cite{gair2011}) and galactic binaries -- for which
various critical parameters like masses, angular positions, and polarisations can be accurately determined
\cite{babak2010}. The output signals will last for years with
very high signal-to-noise ratios.
Observing these signals and accurately estimating the encoded properties of their astrophysical sources places requirements on the level and knowledge of the residual acceleration noise of the instrument reference masses. For eLISA, the acceleration requirement is set at $\unit[3\times10^{-15}]{m\,s^{-2}\,Hz^{-1/2}}$ around $\unit[1]{mHz}$, with an accepted knowledge of roughly 10\% \footnote{to first order, a biased knowledge of 10\% on the instrument noise is equivalent to the same amount of bias on the likelihood employed in GW detection, thus on the astrophysical parameters}.

GW detection relies upon tracing accurately the relative motion
between free-falling test masses (TM) with laser
interferometers. As a GW signal crosses the detector, it
produces a variation of the relative distance, thus a
detectable Doppler shift. The fundamental GW-sensitive element
in such detectors is the Doppler link \cite{vitale2009} between
two free-falling TMs. A recent paper \cite{congedo2013} showed how the time derivative of the measured frequency shift quantifies the integrated curvature along the light beam in terms of relative acceleration, thus the Doppler link effectively behaves as a differential accelerometer. In that paper, a key theoretical aspects were clarified, all in all, the manifestation of the Riemann tensor -- the real covariant observable that can be measured -- as well as the inertial effects. In addition, that work highlighted the possibility to calibrate the data series by subtracting different systematics, measured couplings or actuation forces, including the transients affecting the slow dynamics of low-frequency detectors.

Quantifying the performances of low-frequency detectors in terms of relative acceleration between free falling TMs is a consolidated technique that has also been successfully adopted for the last decade in ground experiments like torsion pendula \cite{carbone2007a,carbone2007b}.

This work shows what are the implications and the key benefits of all these ideas when applied to LISA Pathfinder (LPF)
\cite{vitale2002,antonucci2012,mcnamara2013}, the technology
precursor of eLISA and all future space-based GW detectors. LPF is a reduced version of an interferometric link allocated to fit the size of a spacecraft (SC). Besides a few differences,
LPF shares with the GW detectors a large fraction of the hardware (drag-free and attitude control laws, nm-interferometry, $\mu$N-thrusters, nN-electrostatic actuators, UV-lamp discharging of the TM, etc.) as well as the overall measurement principle. In addition to testing such a challenging technology for GW detection in space, LPF will provide an
accurate model of the instrument acceleration noise, well constrained to within
$\unit[3\times10^{-14}]{m\,s^{-2}\,Hz^{-1/2}}$ around
$\unit[1]{mHz}$ (a factor $\sim10$ relaxed to eLISA).

In recent years, the LPF data analysis \cite{antonucci2011b}
collaboration has been investing effort in two main topics.
Firstly, the estimation of the residual acceleration noise
\cite{ferraioli2010}, and its quantitative analysis
\cite{ferraioli2011b}, is critical for the success of the
mission. Secondly, the estimation of the modelled system
parameters through calibration experiments (see
\cite{congedo2012,nofrarias2012,congedo2011,nofrarias2010}, and
more recently \cite{ferraioli2013}), ensures that disturbances
and systematic errors can be effectively subtracted from the
data. An alternative approach described in a recent work \cite{vitale2014} tries subtracting all disturbances and systematic errors by 
relaxing the a-priori knowledge of the underlying model with
a Bayesian marginalisation over noise, resulting in some marginal posterior, which in principle is found to be equivalent to a re-weighted least squares fitting.

In essence the novel approach contained in this paper is to show how considering acceleration as the baseline dataset and subtract all systematics from it in a model fitting approach can have a few benefits. First, all systematic effects and the actuation forces can be easily subtracted from acceleration. Second, the acceleration dataset itself provides a suitable way to effectively marginalise over the initial conditions that are typically found in displacement and generate long transients in the slow dynamics of low-frequency GW detectors. In standard likelihood estimation, like in GW inference, the focus is typically on the signals, while the noise is assumed be known or mildly changing. For example, in Ref.\ \cite{cornish2007,adams2010}, the authors discriminates between the GW background and the instrument noise allowing the nose floor as a free parameter of the fit. The change of paradigm in this work consists in the following: the signals are now the systematic effects, typically known to within some amplitude or time shift, but the noise is the target of the measurement. 
Working in the acceleration domain instead of displacement turns out to be a convenient way to marginalise over initial conditions and coherently subtract for systematic effects and the actuation forces. Additionally, it is still possible to marginalise over all linear parameters, like in the $\mathcal{F}$-statistic \cite{jaranowski1998}, and over the noise itself, like in Ref.\ \cite{rover2011a,rover2011b,vitale2014}, ultimately ending up with a new statistic whose parameter space is drastically reduced thanks to this multiple marginalisation.

The paper is structured as follows.
In Section \ref{sect:ifo_link} we review a simplified version of the interferometric link for a space-based GW detector, like eLISA, which is useful to introduce the idea of calibrating the data series by coherently subtracting the actuation forces and different systematics induced by the detector itself. This is straightforward if the displacement series is ``filtered'' into equivalent acceleration. The expected GW signal contained in it can be calculated accordingly and the eLISA-like unequal arm Michelson interferometer can also be implemented in terms of acceleration. The similarities and the differences between LPF and the single eLISA arm are finally discussed.
In Section \ref{sect:transients} the transients in displacement data are shown to be nominally suppressed in the equivalent acceleration series, including the extent to which this can be achieved.
In Section \ref{sect:subtraction} we see how the optimal subtraction of asynchronous signals suits perfectly well the need of calibrating acceleration data. This leads to a new statistic that marginalises over all linear parameters, including noise.
In Section \ref{sect:results} we review the application to the LPF data analysis and analysis of a realistic simulator data, make the comparison between different statistics applied to the same dataset and show the agreement between residual acceleration and expected noise level.
Conclusions and the impact that this procedure may have to the analysis of future GW detectors are finally discussed.

\section{The test-mass-to-test-mass interferometric link} \label{sect:ifo_link}

The interferometric link of a space-based GW detectors can be described  
as a sequence of three 
independent interferometric measurements: (\textit{i}) the
motion of the host spacecraft (SC$_1$) relative to a cubic
Gold-Platinum TM$_1$ acting as the reference local free-falling
body; (\textit{ii}) the motion of the second spacecraft SC$_2$
relative to SC$_1$ (at distance of at least
$\unit[1\times10^6]{km}$ equivalent to the flight time of a photon of $T\sim\unit[3]{s}$); (\textit{iii}) the motion of SC$_2$
relative to the second TM$_2$ acting as the reference local
free-falling body. Each SC is forced by $\mu$N-thrust actuators
to follow the reference local TM and compensate for force
disturbances affecting the measurement particularly around
$\unit[1]{mHz}$ (drag-free actuation). In absence of lateral/angular motion of both the SCs and all the optical benches, the three light beams can be assumed be aligned along the same measurement axis $x$ -- the GW-sensitive axis. Also, the proper times of the two free-falling TMs' reference frames is assumed be $\text{d}t_1\simeq\text{d}t_2$, apart a time delay between the two, $t_2=t_1-T$. All the calculation can be done in the local free-falling reference frame of TM$_1$ and $t\equiv t_1$, so 
$x_1(t)$ is the instantaneous position of SC$_1$ relative to
TM$_1$,
$x_2(t)$ is the instantaneous position of SC$_2$ relative to TM$_2$ (but measured at the time of TM$_1$),
$X_{12}(t)$ is the instantaneous position of SC$_2$ relative to SC$_1$,
then the
first-order approximated collinear dynamics of the 3 bodies 
in the limit
of small local motions $x_1$ and $x_2$ is
\begin{subequations}\label{eq:lisa_arm}
\begin{align}
a_1(t) & = \left[\frac{\text{d}^2}{\text{d}t^2} + (1+\tilde{m}_1)\omega_1^2\right]x_1(t) + f_{\text{df},1}(t) , \label{eq:lisa_arm_a} \\
a_2(t) & = \left[\frac{\text{d}^2}{\text{d}t^2} + (1+\tilde{m}_2)\omega_2^2\right]x_2(t) + f_{\text{df},2}(t) , \label{eq:lisa_arm_b} \\
A_{12}(t) & = \frac{\text{d}^2X_{12}(t)}{\text{d}t^2} + \tilde{m}_1\,\omega_1^2\,x_1(t) - \tilde{m}_2\,\omega_2^2\,x_2(t-T) \nonumber \\
& \quad + f_{\text{df},1}(t) - f_{\text{df},2}(t-T) .
\label{eq:lisa_arm_c}
\end{align}
\end{subequations}
$a_1$ and $a_2$ are now the residual
accelerations of SC$_1$ relative to TM$_1$ and SC$_2$ relative
to TM$_2$. $A_{12}$ is the residual acceleration of SC$_2$
relative to SC$_1$,
the former being shifted to the reference time of TM$_1$.
$f_{\text{df},1}$ and $f_{\text{df},2}$ are
the drag-free actuation forces on the two SCs.
$\tilde{m}_1=m_1/m_{\text{SC}_1}$ is a mass normalised to the
SC mass, thus each term multiplied by this factor represents a back-reaction force. $\omega_1^2$
and $\omega_2^2$ ($\sim\unit[-1\times10^{-6}]{s^{-2}}$) are residual (unstable) spring-like
couplings modelling residual force gradients between each TM and the host
SC.

Of course, Eqs.\;\eqref{eq:lisa_arm_a} and
\eqref{eq:lisa_arm_b} describe the two local motions and how
each TM couples locally with the host SC. Instead,
Eq.\;\eqref{eq:lisa_arm_c} describes the SC-SC motion that
contains both the drag-free thrust actuation and the GW signal we are interested in.
Unfortunately, as the drag-free actuation is 
necessary for system stabilisation and the compensation of the radiation pressure acting on the SCs, it inevitably injects actuation noise into the system itself and
makes the extraction of GWs directly from Eq.\;\eqref{eq:lisa_arm_c}
extremely difficult. What really matters for the GW detection
is actually the differential time-delayed motion of the two (almost)
free-falling TMs defined by
\begin{equation}\label{eq:lisa_link}
x_{12}(t)=x_2(t-T)-x_1(t)+X_{12}(t) ,
\end{equation}
that, with the help of Eqs.\;\eqref{eq:lisa_arm}, satisfies the following equation of motion
\begin{equation}\label{eq:lisa_link_acc}
a_{12}(t) = \frac{\text{d}^2x_{12}(t)}{\text{d}t^2} + \omega_2^2\,x_2(t-T) - \omega_1^2\,x_1(t) ,
\end{equation}
where $a_{12}(t) = a_2(t-T)-a_1(t)+A_{12}(t)$ is the calibrated residual time-delayed differential acceleration between free-falling TMs. As the drag-free actuation
forces, $f_{\text{df},1}$ and $f_{\text{df},2}$, and the terms
proportional to $\tilde{m}_1$ or $\tilde{m}_2$ are all internal
action-reaction forces, they clearly disappear in the above
equation. This explains why Eq.\;\eqref{eq:lisa_link} will be
nominally insensitive to the drag-free actuation. In addition, on
the right side of Eq.\;\eqref{eq:lisa_link_acc}, $\omega_1^2$ and $\omega_2^2$
must be subtracted from the second derivative in order to
provide an estimate of the residual acceleration between the
two TMs, cleaned by the two local couplings. All in all, Eq.\;\eqref{eq:lisa_link_acc} shows it seems vital to describe the
sensed motion in terms of equivalent input acceleration in order to subtract the couplings and systematics that would otherwise affect the displacement output in Eq.\;\eqref{eq:lisa_link}.

For the computation of the GW contribution to the acceleration in Eq.\;\eqref{eq:lisa_link_acc} we shall adopt the formalism of Ref.\;\cite{vinet2013} and consider a GW signal coming from the direction $-\hat{w}$ to the detector line of sight $\vec{x}_{12}$. We shall consider the reference frame of TM$_1$ receiving a light ray from TM$_2$ after $T$ seconds. In the limit of small perturbations and small velocities, the GW signal $h_{ij}$ will induce a frequency shift onto the detected laser beam and an equivalent differential acceleration
that will show up in the calibrated ``filtered'' time-series $a_{12}$ as the following signal
\begin{equation}\label{eq:lisa_link_acc_h}
a_{12,h}(t)=
\frac{c}{2(1-\hat{w}\cdot\hat{n})}
\frac{\text{d}}{\text{d}t}\left[H(t-T)-H(t-\hat{w}\cdot\hat{n}\,T)\right],
\end{equation}
where $H=h_{ij}n^i n^j=h_+\xi_+ +h_\times\xi_\times$ ($\xi_+$ and $\xi_\times$ are the so-called antenna patterns) and $n^i\equiv \vec{x}_{12}/cT$. In the long-wavelength limit, the equation further simplifies to
\begin{equation}\label{eq:lisa_link_acc_h_2}
a_{12,h}(t)\simeq-\frac{c\,T}{2} \frac{\text{d}^2 H(t)}{\text{dt}^2},
\end{equation}
which shows that the differential acceleration induced by the GW signal onto the differential motion in Eq.\,\eqref{eq:lisa_link} is effectively proportional to the second derivative of the perturbation itself.

\begin{figure}[H]
\centering
\hspace*{-5pt}\includegraphics[width=0.5\columnwidth]{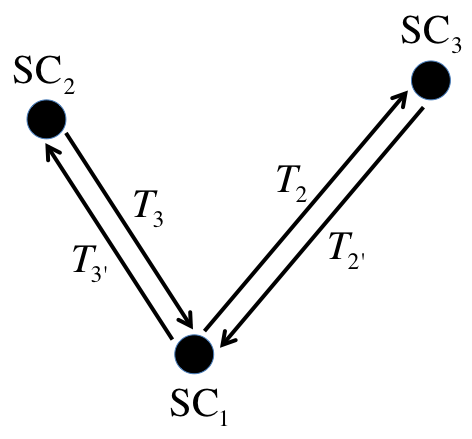}
\caption{\footnotesize{TDI configuration in eLISA with one ``mother'' SC$_1$ hosting two TMs and two ``daughter'' SCs. The SCs are labelled clockwise, while the arms are labelled counterclockwise; the light delays are labelled with primed delays clockwise and unprimed delays counterclockwise. The unequal arm interferometer $\mathit{X}_a$ [see Eq.\;\eqref{eq:tdi}] is physically equivalent to the difference of a synthetic light beam originating from SC$_1$ and bouncing first off SC$_2$ and then off SC$_3$ and another one bouncing first off SC$_3$ and then off SC$_2$.}}
\label{fig:tdi}
\end{figure}

The LPF system is effectively a reduced-size
version of an interferometric link with $X_{12}=0$ and $T=0$ in
Eq.\;\eqref{eq:lisa_link}. Thus LPF will not characterise other peculiar effects like the time delays between different SCs, Doppler shifts due to their differential velocities, shot noise produced by the $\sim\unit[100]{pW}$ light collected by the distant SC, pointing jitter produced by the differential angular motion between SCs and, of course, clock noise and laser frequency noise that are correctly addressed by time delay interferometry (TDI)\cite{tinto2014,tinto2005}. The purpose of TDI is in fact to mitigate the effect of laser frequency noise by forming combinations of the individual Doppler link measurements that suppress laser frequency noise relative to the GW signal. The TDI observables are typically described as combinations of Doppler frequency or phase measurements, which correspond to velocity and position measurements, respectively. The same combinations can be made with acceleration [Doppler frequency derivative, as in Eqs.\;\eqref{eq:lisa_link_acc_h} and \eqref{eq:lisa_link_acc_h_2}] with identical suppression of laser frequency noise. For example, the so-called 1st-generation or ``position-correcting'' Michelson-X TDI variable can be written in terms of the acceleration terms listed above as
\begin{equation}\label{eq:tdi}
\begin{split}
\mathit{X}_a & = \left[(a_{13}+a_{31,2})+(a_{12}+a_{21,3'})_{,22'}\right] \\ 
\quad & -\left[(a_{12}+a_{21,3'})+(a_{13}+a_{31,2})_{,33'}\right],
\end{split}
\end{equation}
where SC$_1$, the so-called ``mother spacecraft", is placed at the vertex of a triangular unequal-arm Michelson interferometer, hosting two free-falling TMs, TM$_1$ and TM$_{1'}$, and two lasers labelled in the same way. TM$_1$ faces TM$_2$ in SC$_2$ and TM$_{1'}$ faces TM$_3$ in SC$_3$, the so-called ``daughter spacecrafts", each containing a laser, again labelled in the same way. Additionally, a time delay is denoted as in $a_{12,3}\equiv a_{12}(t-T_{3})$. See Fig.\;\ref{fig:tdi} for reference. With the same convention of TDI, the SCs are labelled clockwise, while the arms are labelled counterclockwise; the time delays for light beams propagating clockwise (counterclockwise) are labelled with primed (unprimed) indexes. Eq.\;\eqref{eq:tdi} is a combination of four differential acceleration streams, physically equivalent to the difference of a synthetic light beam originating from SC$_1$ and bouncing first off SC$_2$ and then off SC$_3$ and another one bouncing first off SC$_3$ and then off SC$_2$. Although in its simplified form not accounting for the arm flexing and not explicitly including the phase noise coming from the optical benches -- the same substitutions can also be made for the more complex 2nd-generation or ``velocity-correcting" TDI variables -- it shows indeed that it is even possible to develop TDI in terms of accelerations instead of displacements, and build up the unequal-arm Michelson combination $\mathit{X}_a$, while compensating for laser phase noise, but also subtracting actuation forces and detector systematics in each single acceleration term $a_{ij}$ -- whose characterisation with LPF is in fact the goal of this paper.

The LPF dynamics \cite{armano2009} shows up one key difference to eLISA: the interferometric link is hosted within one single SC and, as such, only one TM can be in free fall along the measurement axis $x$. This will act as the reference local free-falling body and the SC will be forced to follow that through drag-free actuation. In the meantime the other TM will be forced to follow the reference TM through electrostatic actuation. Thus the dynamics along the sensitive
axis $x$ can be described in terms of two degrees of freedom,
the relative motion of the SC to the reference TM
($x_1$) and the relative motion between the two TMs
($x_{12}$). Additionally: (\textit{i})
residual spring-like couplings $\omega_1^2$, $\omega_2^2$ (to
current best knowledge from ground measurements and models,
$\sim\unit[-1\times10^{-6}]{s^{-2}}$), and their difference
$\omega_{12}^2=\omega_2^2-\omega_1^2$; (\textit{ii}) the
gravitational gradient $\Gamma_{12}$ between the TMs
($\sim\unit[1\times10^{-9}]{s^{-2}}$); (\textit{iii}) the
drag-free actuation $A_\text{df}f_\text{df}$ on the SC (with
actuation gain $A_\text{df}$); (\textit{iv}) the electrostatic
suspension actuation $A_\text{sus}f_\text{sus}$ on TM$_2$
(with actuation gain $A_\text{sus}$).
The adopted model is as follows
\begin{subequations}\label{eq:lpf_arm}
\begin{align}
a_1(t) & = \left[\frac{\text{d}^2}{\text{d}t^2} + (1+\tilde{m}_1)\omega_1^2 + \tilde{m}_2\,\omega_2^2\right]x_1(t) \nonumber \\
& \quad + \left(\Gamma_{12}+\tilde{m}_2\,\omega_2^2\right)x_{12}(t) \nonumber \\
& \quad + A_\text{df}\,f_\text{df} - \tilde{m}_2\,A_\text{sus}\,f_\text{sus}(t) , \\
a_{12}(t) & = \left[\frac{\text{d}^2}{\text{d}t^2} + \omega_2^2-2\Gamma_{12}\right]x_{12}(t) \nonumber \\
& \quad - \omega_{12}^2\,x_1(t) + A_\text{sus}\,f_\text{sus}(t) .\label{eq:lpf_arm_b}
\end{align}
\end{subequations}
The observed motion is typically slightly different from the true motion, $x_1$ and $x_{12}$. Such a difference accounts for $\sim1\times10^{-4}$ of $x_1$-motion leaking into the baseline differential measurement$x_{12}$.

In the next sections we shall see how describing the detector dynamics in terms of differential accelerations, $a_{12}$, instead of displacement,$x_{12}$, can help mitigate or subtract spurious effects due to the couplings with the SC and the actuation. 

\section{Filtering to acceleration data and mitigating for transients} \label{sect:transients}

The interferometric link of eLISA and its miniature version implemented on LPF are two examples of closed-loop feedback control systems, in which the differential acceleration or force per unit mass, $a_{12}(t)$, is the input, and the observed displacement $x_{12}(t)$ is the output. The output, together with other control signals, are fed back as actuation force (per unit mass), $f_\text{act}(t)\equiv f_\text{act}[x_{12}(t),...]$, such that it compensates for disturbances and ensures the linearity of the system within the operating range. The evolution of the system can then be described in terms of a second-order linear inhomogeneous differential equation \cite{congedo2012,antonucci2011a,congedothesis2012}
\begin{equation}\label{eq:diff}
\Delta\,x_{12}(t)=a_{12}(t) .
\end{equation}
The action of $\Delta$ is to physically compute the second derivatives of the observed differential motion $x_{12}$, apply some coupling coefficients, subtract measured signals acting as systematic effects we are interested in eliminating, and subtract the applied actuation forces. In general, there exists a steady-state solution,
$x_{12}^\text{steady}(t)$, depending on the driving inputs
$a_{12}(t)$, and a transient solution, $x_{12}^0(t)$,
depending on the initial conditions $x_{12}^0\equiv x_{12}(t_0)$ and
$\dot{x}_{12}^0\equiv\dot{x}_{12}(t_0)$. $x_{12}^0(t)$ is in fact a particular
solution of the homogeneous equation
$\Delta\,x_{12}(t)=0$, for which in principle
the kernel space might be non trivial, i.e.\ there exists
a set of coefficients $\sigma_j$, so that
$x_{12}^0(t)=\sigma_j\,\phi_j(t)$ (implicit summation over repeated indexes), where all the $\phi_j(t)$ satisfy the homogeneous
equation. By definition, it follows that by applying
$\Delta$ to the general solution, $x_{12}^\text{steady}(t)+x_{12}^0(t)$, it suppresses the transients, 
\begin{equation}
\Delta\left[x_{12}^\text{steady}(t)+\sigma_j\,\phi_j(t)\right]\equiv a_{12}(t) .
\end{equation}
In other words, the free evolution of the system described by initial conditions $\sigma_j$ and transient functions $\phi_j(t)$ will give no contribution to the acceleration $a_{12}(t)$. This will only be driven by the steady state only.

A necessary condition for a system
to be linear and time invariant (hence causal) is to have all
null initial conditions \cite{oppenheim1996}. As such, the
system is completely determined by its impulse response. On the
contrary, modelling the system
in frequency domain with a set of input-output transfer
functions in presence of non-null initial conditions,
makes the free evolution of the system undetermined. We are now faced with a dilemma: either follow the system dynamics in the displacement space or acceleration space. However the two approaches would be totally equivalent if all initial conditions were exactly zero. In fact all dynamical equations can be solved in displacements or accelerations with a univocal relationship, as well as any likelihood would be the same regardless of the space we choose to work in. On the contrary, if any initial condition were different from zero, then (\textit{i}) the canonical transfer function approach would fail, except if one uses 1-sided Laplace transforms, but paying the price of much more complexity; (\textit{ii}) in the displacement space we would need to solve the system of differential equations and any likelihood estimation would require many more free parameters (including the initial conditions), thus we end up again with much more complexity than really required. As the initial conditions are essentially nuisance parameters we are not interested in, it is much more valuable to follow the dynamics in the acceleration space where the transients are automatically mitigated and the model can be written as an (almost) linear combination of time series (compare this with the integrals one may have in the displacement domain).

The differential operator $\Delta$ filtering the sensed motion into equivalent input acceleration can be easily identified in both the equations of motion for eLISA [Eq.\;\eqref{eq:lisa_link_acc}] and LPF [Eq.\;\eqref{eq:lpf_arm}]. It contains second derivatives as well as the parameters modelling various couplings. Effectively, any coupling (e.g.\ $\omega_1^2$ or $\omega_{12}^2$) enters into $\Delta$ as
a coefficient, which must be accurately determined in advance with calibration experiments before the
operator can be actually used to make an unbiased estimate of the residual acceleration. The linearity of the operator is ensured as long as the system dynamics itself is linear. This approximation holds true if the control laws are properly set up such that in absence of external forces the system remains stable; it even remains stable if an external force is applied within certain allowed range.

Filtering the displacement data into acceleration has the clear advantage that the actuation forces [see Eq.\;\eqref{eq:lpf_arm}] can be retrieved from the telemetry and subtracted from acceleration. Evidently, working with displacement data the actuation forces cannot be easily modelled without proper transfer functions. Using directly the equations of motion provides a clear advantage over inverting the equations themselves and fitting them to displacement data. However, inaccuracies in the measured parameters, say $\delta\Delta$, impact on the knowledge
of the residual acceleration as shown, for instance, in
Ref.\;\cite{congedo2012}, with the systematic error $\delta\Delta\,x_{12}(t)$. In the same way, such inaccuracies affect the suppression of
transients, which lasts considerably long compared to the typical duration of
a LPF experiment \footnote{For instance in LPF, the typical duration of
an identification experiment is $\sim\unit[3]{h}$, whereas
transients induced by changes of the system configuration might
last for $\sim\unit[1]{h}$.} with an unsuppressed systematic error $\delta\Delta\,\phi_j(t)$. Again, ``tuning up'' the operator is critical to make an unbiased measurement of acceleration free of unwanted effects. 

Given the slow dynamics of $\unit{mHz}$ detectors, which are quite susceptible to tiny changes in the system configuration or non-stationary noise fluctuations, they will be inevitably dominated by long transients. Once properly calibrated, only filtering the detector output into acceleration can mitigate such an effect in an effective marginalisation over initial conditions, with the additional benefit that any other systematic signal can be subtracted as well.

\section{Optimal subtraction of asynchronous signals}
\label{sect:subtraction}

In this section we shall describe a technique that implements the optimal subtraction of known asynchronous signals. Suppose $x$ is the dataset. We don't particularly focus on displacement or acceleration data for the moment, but it is clear that the equations of motion can be written as a linear combination of time-shifted signals $s_i$, plus some noise $n$
\begin{equation}\label{eq:coherent_subtraction}
x(t) = \theta_i s_i(t-\tau_i) + n(t) ,
\end{equation}
with implicit summation over repeated indexes. $\vec{\theta}=\theta_i$ are amplitude parameters and $\vec{\tau}=\tau_i$ are shift parameters we aim at measuring, such that all measured $s_i$ can be coherently subtracted from $x$. The result is an estimate of $n$, a zero-mean Gaussian process with power spectral density (PSD) $S_\text{n}$. For the moment, let's assume $S_\text{n}$ is known and does not depends on parameters. Here the change of paradigm compared to standard likelihood estimation, as for instance in GW inference, is that we are not interested in extracting $s_i$ from noisy data, rather we are interested in measuring $\vec{\theta}$ and $\vec{\tau}$ such that we can coherently subtract $s_i$ and estimate $n$.

As customary in GW inference, we define the likelihood in terms of the noise-weighted inner product in frequency domain
\begin{equation} 
(x|y)=\int \tilde{x}(\omega)^*S_\text{n}(\omega)^{-1}\tilde{y}(\omega)\,\text{d}\omega ,
\end{equation} 
and $S_\text{n}$ for a discrete time-series can safely be assumed diagonal. With Gaussian noise, the log-likelihood of the problem is given by
\begin{equation}\label{eq:log-likl}
-2\log\mathcal{L}(x|\vec{\theta},\vec{\tau}) = (x-\theta_i s_i(\tau_i)|x-\theta_i s_i(\tau_i)) .
\end{equation}
Now, the likelihood can be optimised analytically with respect to the linear parameters $\vec{\theta}$ and the solution is
\begin{equation}\label{eq:fisher}
\theta_i = (F^{-1})_{ij}(x|s_j) ,
\end{equation}
where $F_{ij}$ is the Fisher matrix with respect to $\vec{\theta}$
\begin{equation}
F_{ij} = (s_i|s_j) .
\end{equation}
As $\vec{\theta}$ are now effectively functions of $\vec{\tau}$, the latter will be the only free parameters of the fit and the dimensionality of the problem is dramatically reduced. For instance, if we want to fit $N_\text{s}$ signals, each one allowed to vary both in terms of amplitudes and shifts, the dimensionality of the problem for standard likelihood estimation would be $2N_\text{s}$, but implementing this analytical optimisation the number becomes just $N_s$, making the exploration of the parameter space much easier. This extremal likelihood, obtained by plugging Eq.\;\eqref{eq:fisher} into Eq.\;\eqref{eq:log-likl} and in widespread use throughout the whole GW community both on ground and in space, is known as the $\mathcal{F}$-statistic\cite{jaranowski1998} \footnote{Contrary to what was proposed by the authors and to assume here the probabilistic point of view, in this work the statistic is in fact denoted with the same symbol as the distribution itself, just to ease the notation. A change between the two implies only the application of the log-function.}. However, it is worth noting that this statistic is just equivalent to marginalising $\mathcal{L}$ over $\vec{\theta}$ assuming a flat and infinite uninformative prior $p(\vec{\theta})$
\begin{equation}\label{eq:f-statistic}
\begin{split}
\mathcal{F}&(x|\vec{\tau})  = \int \mathcal{L}(x|\vec{\theta},\vec{\tau})p(\vec{\theta})\,\text{d}\vec{\theta} \\
& = \exp\left[\frac{1}{2}(F^{-1})_{ij}(x|s_i(\tau_i))(x|s_i(\tau_j))-(x|x)\right] ,
\end{split}
\end{equation}
where $(x|x)$ is just a constant. This result that can be easily obtained through a simple Gaussian integration.

Having found the connection between the standard likelihood and the $\mathcal{F}$-statistic, we shall now extend that to unknown $S_\text{n}$ as in Ref.\ \cite{vitale2014}. In that work, the authors addressed the problem of estimating the noise parameters in a Bayesian approach by relaxing the knowledge on $S_\text{n}$. This is done by marginalising the likelihood over a flat prior as a function of either $\log{S_\text{n}}$ or some other slow function of $S_\text{n}$. The result was that the posterior is given by the sum of squared log-residuals, referred to as the $\Lambda$-statistic, instead of the sum of squares as in standard likelihood. The marginalisation over $S_\text{n}$ effectively corresponds to a change in definition of the noise-weighted inner product to the following
\begin{equation} 
(x,y)_{\log}=\int \log\left[{\tilde{x}(\omega)}^*\tilde{y}(\omega)\right]\text{d}\omega .
\end{equation}
and the marginalised likelihood becomes
\begin{equation}\label{eq:log-likl}
-2\log\Lambda(x|\vec{\theta},\vec{\tau}) = (x-\theta_i s_i(\tau_i),x-\theta_i s_i(\tau_i))_{\log} .
\end{equation}
However, it is clear that a good fraction of all parameters involved in the estimation are linear indeed, so it is still possible to marginalise them out before integrating over the noise prior. This work extends the $\mathcal{F}$-statistic to the case of unknown coloured noise and the result is a new statistic. The calculation is straightforward.
We wish to marginalise $\mathcal{L}(x|\vec{\theta},\vec{\tau},S_\text{n})$ over a joint flat prior of all $\vec{\theta}$ and $\log{S_\text{n}}$ \footnote{Alternatively, we could choose some slow function of $S_\text{n}$ without affecting the results.}, apply a prior on $\vec{\tau}$, and finally obtain the full marginalised posterior
\begin{align}
\Phi(\vec{\tau}|x) & = 
\int_{\vec{\theta}}\int_{S_n} \mathcal{L}(x|\vec{\theta},\vec{\tau},S_\text{n})p(\vec{\theta})p(\vec{\tau})p(S_\text{n})\,\text{d}\vec{\theta}\,\text{d}S_\text{n} \nonumber \\
& = \int_{S_n}\mathcal{F}(x|\vec{\tau},S_\text{n})p(\vec{\tau})p(S_\text{n})\,\text{d}S_\text{n} \nonumber \\
& = \Lambda^\mathcal{F}(x|\vec{\tau})p(\vec{\tau}) .
\end{align}
The interpretation of the calculation is the following. In the first integral, the standard likelihood is marginalised over the linear parameters and the $\mathcal{F}$-statistic comes out naturally, this corresponds to a linear fit and has the clear advantage that it drastically reduces the dimensionality of the problem. Then in the second integral, we marginalise the $\mathcal{F}$-statistic over the noise, like previously discussed, by computing the sum of log-squared residuals with all linear parameters replaced with their best-fit values -- this is $\Lambda^\mathcal{F}$. The result of this double analytical integration, and the application of a prior on $\vec{\tau}$ -- the only remaining parameters to fit -- takes to a full Bayesian posterior referred to as the $\Phi$-statistic. This correctly takes into account of the (at least in principle) unknown noise shape and marginalise over all linear parameters. 

\section{Results} \label{sect:results}

This section turns to the application of the method to LPF, whose outcome is to ultimately provide an unbiased estimation of acceleration noise, cleaned up by systematic effects and applied actuation forces.  But, first of all, we review how the experiments are actually executed and analysed in LPF.
The LPF dynamics can be understood in details
with calibration experiments by stimulating the system along different degrees of freedom. Then, with an accurate knowledge of the various parameters, the residual acceleration can be measured. The typical identification experiment is a sequential injection of sinusoidal signals (either forces or control signals
between $\unit[1]{mHz}$ and $\unit[50]{mHz}$ with an integer number of cycles) into a particular
degree of freedom. We shall call it the \textit{injection} experiment, in contrast to the \textit{noise-only} experiment where no signal will be injected. The system will thus react to the injections at the output displacement between the two TMs and the
experimental data will be
$x_{12}=x_{12,s}+x_{12,n}$,
where $s$ stands for the signal produced by the
injection and $n$ stands for the instrument
noise we wish to measure. 
By applying the differential operator $\Delta$, as in
Eq.\;\eqref{eq:diff}, directly to the displacement data, the equivalent input differential acceleration can be estimated. So, focusing on $x_{12}$, it is possible to assume that the experimental data relevant for this analysis are actually $a_{12}=\ddot{x}_{12}$. Clearly, the method derived in the preceding section and based on the sequential subtraction of measured signals applies directly to this problem, thanks to its close analogy with the equations of motion themselves. In other words, it is straightforward to apply the optimal statistic and model the equations of motion of both the eLISA link and the LPF system as coherent subtraction of measured, but asynchronous, signals. Therefore, the 
log-likelihood becomes
\begin{equation}\label{eq:log-likl-acc}
-2\log\mathcal{L}(a_{12}|\vec{\theta},\vec{\tau}) = (a_{12}-\theta_i s_i(\tau_i)|a_{12}-\theta_i s_i(\tau_i)) ,
\end{equation}
where the inner product is calculated based on $\hat{S}_{a_{12,\text{n}}}$, which can either estimated from noise-only measurements or directly available as a theoretical model or even marginalised over using the full $\Phi$-posterior. The above likelihood also implements the action of the dynamical $\Delta$ operator, where its parameters, now $\vec{\theta}$ and $\vec{\tau}$, enter into the likelihood calculation. By using the $\Phi$-statistic we incidentally marginalise over all noise parameters, including the unknown PSD we ultimately aim at measuring. It is worth noting that this complete analogy between equations of motion and optimal coherent subtraction of signals is only available if one decides to treat the differential acceleration as the real data domain to work with. On the contrary, signals are not easily subtracted, at least not as easy as a quasi-linear combination like in Eq.\;\eqref{eq:coherent_subtraction}, in displacement domain where, for instance, ones needs proper transfer functions to model the effect of applied forces to displacement. Another side bonus is that, by definition, the $\Delta$ operator implements the system dynamics. So, by calibrating the noise parameters, the free evolution of the system, i.e.\ the response to non-null initial conditions in displacement data, is mitigated in the acceleration domain.

To make the physical meaning of the above noise parameters clearer, we shall now consider LPF in more details: the subtracting signals can be divided in two main categories: linear/angular motion relative to each of the two TMs, plus the angular motion of the SC (in total, 6+6+3=15 signals) and forces/torques applied to all bodies (in total, 18 signals). Following the analogy with the equations of motion, $\theta$-parameters multiplying the observed motion (either linear or angular) are collectively called \textit{stiffness} coefficients, because they are effective spring-like constants coupling the differential acceleration along the optical axis $x$ with any degree of freedom. Instead, $\theta$-parameters multiplying applied forces/torques are collectively called \textit{gain} coefficients, because they are effective gains translating how the commanded forces are actually applied by the actuators to the system and again translated into observed acceleration. Additionally, the $\tau$-parameters, effective time-shifts applied to the signals been subtracted from the acceleration data, are originated within the closed-loop dynamics of LPF where the thrust and electrostatic actuators, the on-board data management unit, the electrical buses, overall account for some fraction of a second depending on each particular signal.

Once all parameters have been measured, an estimate of the differential acceleration noise is given by
\begin{equation}
\hat{a}_{12,n}(t,\vec{\theta},\vec{\tau}) = a_{12}(t)-\theta_i s_i(t-\tau_i)
\equiv \Delta_{\vec{\theta},\vec{\tau}}\,x_{12}(t),
\end{equation}
where we essentially review this operation as the action of the $\Delta$ operator on $x_{12}$ through a second derivative and the coherent subtraction of signals. The operator effectively filters displacement data into a calibrated acceleration dataset.
Therefore the likelihood estimator of Eq.\;\eqref{eq:log-likl-acc} can be reviewed as a two-stages filter:
\begin{itemize}[leftmargin=*]
  \renewcommand{\labelitemi}{-}
  \setlength{\itemsep}{0pt}
  \setlength{\parskip}{0pt}
  \setlength{\parsep}{0pt}
  \item An optimal time-domain filter $\Delta_{\vec{\theta},\vec{\tau}}$ ``filtering'' displacement data into acceleration data. This subtracts for the couplings modelled by $\vec{\theta}$ and $\vec{\tau}$ and mitigate for system transients.
  \item An optional frequency domain filter $\hat{S}^{-1/2}_{a_{12,\text{n}}}$, when known from noise-only experiments, that whitens the acceleration data. This accounts for the very large dynamical range of correlated noise, typical of GW detectors.
\end{itemize}
It is worth noting here that the $\Phi$-statistic presented in this work would skip the second step thanks to the full marginalisation over noise PSD and $\theta$-parameters.

In Table \ref{tab:comparison} we report a comparison between the different estimators discussed in this work and applied to data produced by a realistic 3-dimensional LPF simulator, which has been extensively employed in test campaigns in preparation of the mission. Differential acceleration data have been fit with $s_i=(x_1,x_{12},f_\text{sus})$, i.e.\ the measured differential motion between the two TMs, the relative motion of TM$_1$ with respect to the SC and the commanded electrostatic actuation on TM$_2$. Amplitudes and shifts have been measured with the four estimators by means of a numerical optimiser (conjugate gradient as initial search plus direct search for refinement): standard likelihood ($\mathcal{L}$), 6 free parameters; marginalised likelihood over noise ($\Lambda$), 6 free parameters; marginalised likelihood over linear parameters ($\mathcal{F}$), 3 free parameters; marginalised likelihood over noise and linear parameters ($\Phi$), 3 free parameters. In the first test (columns on the left of each estimator's results), the knowledge of noise has been relaxed, arbitrarily assumed $1\times10^{-26}+10^{-20}f^2\;\unit{m^2\,s^{-4}\,Hz^{-1}}$ where needed for the estimation. In the second test (columns on the right), the noise is measured from a noise-only run, where no signals have been injected to, so ti represents the baseline noise floor. Clearly, all estimates are consistent, within the expected statistical uncertainty ($\sigma_\text{fit}$), but the $\Lambda$ estimator takes many more function evaluations to reach the maximum. Because the $\mathcal{F}$ and $\Phi$ explore the same parameter space, but collapsed to half the dimensionality, they are consequently much more efficient while retaining the required accuracy, also proven by the analysis of residuals that show no systematics. It is expected that this approach should improve any type of sampling, like MCMC, one would employ to explore the parameter space.

\begin{figure}[H]
\centering
\hspace*{-5pt}\includegraphics[width=1.05\columnwidth]{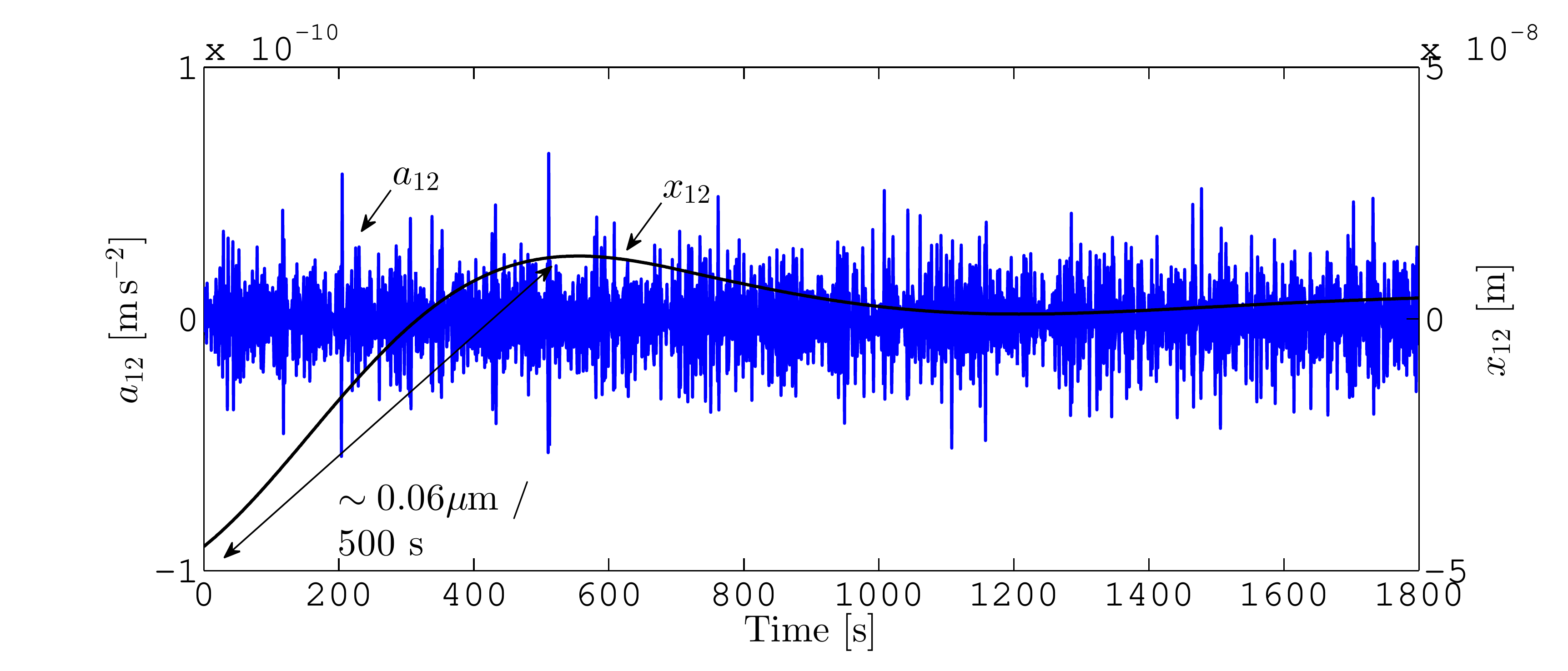}
\caption{\footnotesize{Differential acceleration and displacement time-series for the nominal LPF measurement configuration. The dynamical range is dramatically reduced from displacement to acceleration. The measured acceleration is nominally insensitive to the free evolution of the system. Fitting directly acceleration turns out to be much easier than displacement.}}
\label{fig:acc_transients}
\end{figure}

In another test, the actuation force $f_\text{sus}$ has been fit with $s_i=(x_1,x_{12},a_{12})$ (note that the differential acceleration is now the signal been subtracted). The results is that, with the same model, the fit can proceed analogously with good results because the equations of motion are invariant under recasting the order of the various terms. On the contrary, fitting the differential displacement $x_{12}$ with the same signals   $s_i=(x_1,a_{12},f_\text{sus})$ has not given good success. This may be due to the fact that the transfer function from $f_\text{sus}$ to $x_{12}$ is not trivially modelled by amplitude and shift like with acceleration. Moreover, the displacement signal shows up the presence of long lasting transients that are not easily fit with the simple model that can instead be applied in acceleration domain. Fig.\;\ref{fig:acc_transients} qualitatively shows how the displacement series looks like compared to the acceleration series: while in displacement an initial decay lasting a good fraction of the entire length is observed in the time-series, the acceleration appears much more stable and probably easier to fit. As previously discussed, the real advantage of fitting acceleration is that the system free evolution is nominally suppressed and modelling can be trivially implemented as a quasi linear combinations of signals. 

\begin{table*}
\begin{tabular}{l @{}*{8}{D{.}{.}{5}}@{} c} 
\hline \hline
& \multicolumn{2}{c}{$\mathcal{L}$} & \multicolumn{1}{c}{$\Lambda$} & \multicolumn{2}{c}{$\mathcal{F}$} & \multicolumn{2}{c}{$\Phi$} & \multicolumn{1}{c}{$\sigma_\text{fit}$} & phys.\ param. \\
\hline
$\theta_1$ [$\unit[10^{-6}]{s^{-2}}$] & 0.749 & 0.748 & 0.756 
& 0.749 & 0.748 & 0.749 & 0.748 
& 0.056 & $\left|\omega_1^2\right|$ \\
$\theta_2$ [$\unit[10^{-6}]{s^{-2}}$] & 2.13 & 2.13 & 2.13 
& 2.13 & 2.13 & 2.13 & 2.13 
& 0.20 & $\left|\omega_{12}^2\right|$ \\
$\theta_3$ [$\unit{kg^{-1}}$] & 0.5356 & 0.5356 & 0.5357 
& 0.5356 & 0.5356 & 0.5356 & 0.5356 
& 0.0013 & $A_\text{sus}/m_2$ \\
\hline
$\tau_1$ [s] & 0.153 & 0.096 & -0.070
& 0.153 & 0.096 & -0.052 & -0.054 
& 0.075 & - \\
$\tau_2$ [s] & -0.172 & -0.080 & -0.029 
& -0.172 & -0.080 & -0.087 & -0.081 
& 0.095 & - \\
$\tau_3$ [s] & 0.4923 & 0.4965 & 0.4980 
& 0.4922 & 0.4965 & 0.4965 & 0.4971 
& 0.0024 & - \\
\hline
$n_\text{eval}$ & \multicolumn{1}{c}{641} & \multicolumn{1}{c}{940} & \multicolumn{1}{c}{1283} 
& \multicolumn{1}{c}{246} & \multicolumn{1}{c}{209} & \multicolumn{1}{c}{285} & \multicolumn{1}{c}{305} 
& \multicolumn{1}{c}{-} & - \\
\hline
\hline
\end{tabular}
\caption{\footnotesize{Estimated values with different statistics (see text for details). Two tests are reported (left and right columns for each estimator's results): left, unknown noise shape, arbitrarily assumed to be $1\times10^{-26}+10^{-20}f^2\;\unit{m^2\,s^{-4}\,Hz^{-1}}$; right, known noise shape measured from an independent noise run. The results are consistent with and without knowledge of the underlying noise, but typically $\mathcal{F}$ and $\Phi$ requires less likelihood/posterior evaluations they explores a collapsed parameter space. Expected statistical uncertainty and physical parameters are also included for completeness.}}
\label{tab:comparison}
\end{table*}

\begin{figure*}[htb]
\centering
\includegraphics[width=0.9\textwidth]{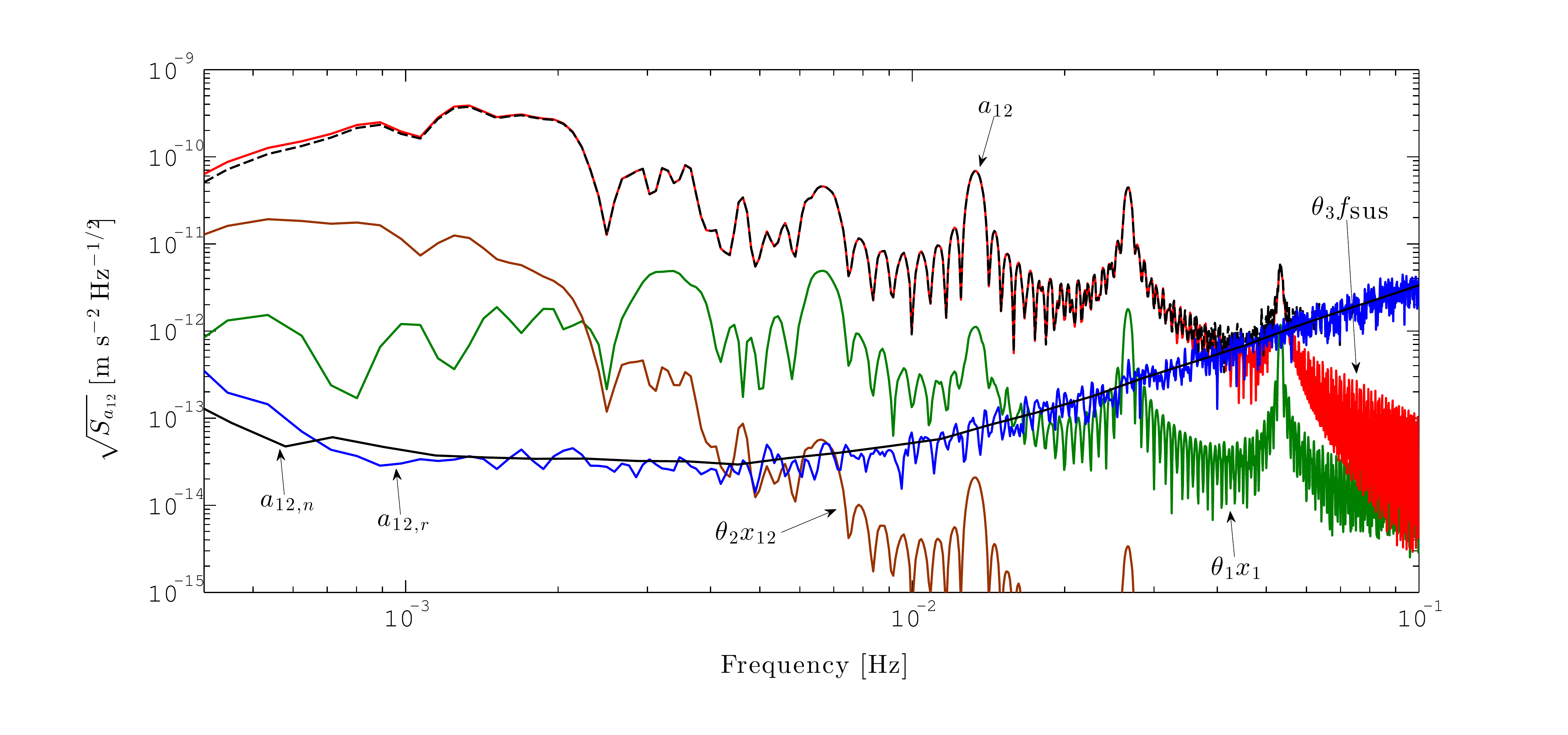}
\vspace*{-20pt}
\caption{\footnotesize{Contributions to a one-day measurement of the residual differential acceleration,
$a_{12,\text{r}}$. In sequence, measured signals corresponding to the electrostatic suspension actuation and the two displacement outputs have been
coherently subtracted from the total acceleration $a_{12}$. All signals injected in the main
frequency band (the peaks between 1 and $\unit[50]{mHz}$) are suppressed, thus allowing the accurate recovery of the noise floor, $a_{12,\text{n}}$ down to a fraction of mHz.}}
\label{fig:acc_proj}
\end{figure*}

In terms of differential acceleration, Fig.\;\ref{fig:acc_proj} shows the contribution of the signals $s_i=(x_1,x_{12},f_\text{sus})$ to the total acceleration $a_{12}$. The residual acceleration $a_{12,r}\equiv a_{12}-\theta_i s_i(\tau_i)$, the parameters being measured with the method presented in this paper, is also shown for comparison and is the final estimate of the true acceleration noise. The estimated PSD
\footnote{The PSD is computed with the Welch overlap method
\cite{welch1967}. The time-series is split in 4 segments, with 50\%
overlap, linear de-trended, and a 4-term Blackman-Harris window \cite{harris1978} is finally applied to each segment. The PSD is computed on each segment and averaged out. The first 4 frequency bins are thrown away because they are affected by the window
\cite{ferraioli2010}.}
of each term is finally compared to the acceleration noise, $a_{12,n}$, measured on an independent noise-only run. The optimal subtraction allows an accurate estimate of each contributing term. The largest is, of course, the electrostatic suspension actuation on TM$_2$, which will be absent in eLISA. An order of magnitude below come the last two terms proportional to $x_{1}$ and $x_{12}$. The proportionality constants, $\theta_1\equiv|\omega_1^2|$ and $\theta_2\equiv|\omega_{12}^2|$, are effective spring-like constants coupling the relative motion of the two TMs. These contributions are expected affect the low-frequency noise of GW detectors. Finally, the residual acceleration is recovered -- no residual peaks corresponding to the injected signals are identified -- and it is well in agreement with the reference noise measurement down to mHz-frequency. By coherently subtracting the signals, which acts as systematic effects, the noise has been effectively reduced by orders of magnitude. Therefore, calibrating those parameters ensures the proper calibration of the acceleration noise, free from systematic errors. 

\section{Concluding remarks}

This paper has put a general methodology framework with which assess the low-frequency acceleration noise between free-falling test masses -- the key measurement element of future GW detectors.

First, an interferometric link of space-based GW detectors like eLISA has been considered and this intuitively shows how the displacement data can be ``filtered'' into acceleration. By doing so, the drag-free actuation is subtracted and different systematic effects can be calibrated as well. The expected equivalent acceleration GW signal has been calculated. Ultimately this shows that acceleration-calibrated data can be used for GW inference, nominally free of detector systematics. Those ideas can also be included into TDI, particularly important for laser frequency noise cancellation, and the X-Michelson combination has been presented to include calibrated differential acceleration data.

Second, as the eLISA link and its miniature version implemented in LPF are examples of closed-loop feedback dynamical systems, the solution of the equations of motions are typically described in terms of a steady-state and a free evolution, i.e.\ the transient solution depending on the initial conditions. In fact, long transients in displacement data are expected for the the low-frequency detectors. On the contrary, acceleration should be nominally insensitive to that effect, as the free evolution is, by definition, suppressed if one chooses to use equivalent acceleration as the baseline dataset. However, being the dynamical operator parameter dependent, it is crucial to assess those parameters in a robust way and ensure an unbiased measurement of acceleration noise.

Third, thanks to the strong similarity between equations of motion and coherent subtraction of known signals in an (possibly) unknown noise, a general and optimal method has been presented. This is a generalisation of previous work. The $\mathcal{F}$-statistic -- well-known among the GW community -- is found to be equivalent to an analytical marginalisation over linear parameters assuming a flat uninformative prior. In other work, a different likelihood, the sum of log-squared residuals, can be obtained by marginalising over the noise PSD and, of course, is useful when noise is not fully known. In this work both techniques has been employed at the same time, thus allowing to marginalise over: (\textit{i}) all the linear parameters; (\textit{ii}) the noise PSD; (\textit{iii}) the system free evolution if acceleration is adopted as the baseline dataset where to perform any likelihood calculation.

Finally, the method has been applied to LPF simulator data. The simulator is a great opportunity to test data analysis ideas in a realistic simulation-like environment. In previous approaches, parameter estimation and measurement of acceleration noise were thought to operate alternately in sequence. Now, the coherent subtraction of asynchronous signals devised in this paper, and its close analogy with the equations of motion, suits perfectly well the need of calibrating LPF data with just a single technique by subtracting the actuation forces and other systematic signals. The method has been used in order to make a comparison between different statistics on the same data. Besides all providing consistent results, the new statistic presented in this work is much more efficient as it explores a reduced-size parameter space, thanks to the multiple marginalisation over linear parameters and noise. Also, it allows to recover the reference noise floor with an accurate subtraction of the various contributions, thus showing the unbiasedness of the measured residual acceleration.

Been very general, the method can be applied, as it is, to any LPF configurations/experiments: what is required is to properly choose the signals to subtract from the acceleration data and this depends on the particular configuration. Thanks to the marginalisation over linear parameters, it is expected that the new statistic should dramatically improve the exploration of the parameter space, in particular when the number of signals to subtract is large. It is in fact well-known that the dimensionality can become an issue for MCMC sampling techniques (an application of which to LPF data can be found, for instance, in Ref.\ \cite{karnesis2014}): the marginalisation over many linear parameters can be the turning point toward an accurate noise measurement in a large-dimensional problem.

The method -- once properly generalised to work with TDI [see for instance the first-generation unequal-arm X-Michelson combination in Eq.\;\eqref{eq:tdi}] -- might also be applied to compensate for laser phase noise, coherently subtract for actuation forces and detector systematics, and mitigate for system transients in eLISA-like data. This is far beyond the scope of this paper, but thanks to its flexibility it is reasonable to expect positive results even in that case. This investigation might be the focus of future work.

%\appendix
%\input{./sections/appendix}

\begin{acknowledgments}
The author acknowledges support from the Beecroft Institute for Particle Astrophysics and Cosmology. The simulator data used in this work were provided by the LISA Pathfinder collaboration. Part of the analysis was done using the LTPDA Toolbox \cite{ltpda}. The author thanks the entire LISA Pathfinder collaboration, and in particular L.\ Ferraioli, M.\ Hewitson, M.\ Hueller, N.\ Karnesis, E.\ Plagnol, J.\ I. Thorpe, S.\ Vitale, W.\ J.\ Weber, for fruitful interaction with regarding this work. The author also thanks M.\ Tinto and M.\ Vallisneri for useful clarification on time delay interferometry.
\end{acknowledgments}

\bibliography{references}

\end{document}